\documentclass[aps, pra , reprint,superscriptaddress]{revtex4-2}

\usepackage[utf8]{inputenc}
\usepackage{graphicx}
 \usepackage{subcaption}
\usepackage{amsfonts}
\usepackage{csquotes}
\usepackage{multirow} 
\usepackage{amsmath}
\usepackage{amsthm}
\usepackage{hyperref}
\usepackage{braket}
\usepackage{xcolor}
\usepackage{bbm}
\usepackage{comment}
\usepackage{tikz}

\usepackage{ragged2e}

\AtBeginDocument{\let\raggedright\justifying}

\newcommand{\phase}[1]{\text{ang}[#1]}
 
\usepackage[sort&compress]{natbib}
\usepackage{wrapfig}
\begin{document}

\title{{Scalable Quantum Interference from Indistinguishable Quantum Dots}}
 
\author{Sheena Shaji}
\altaffiliation{These authors contributed equally to this work.}
\affiliation{Institute of Photonics and Quantum Sciences, Heriot-Watt University, Edinburgh EH14 4AS, United Kingdom}

\author{Suraj Goel}
\altaffiliation{These authors contributed equally to this work.}
\affiliation{Institute of Photonics and Quantum Sciences, Heriot-Watt University, Edinburgh EH14 4AS, United Kingdom}

\author{Julian Wiercinski}
\altaffiliation{These authors contributed equally to this work.}
\affiliation{Institute of Photonics and Quantum Sciences, Heriot-Watt University, Edinburgh EH14 4AS, United Kingdom}

\author{Frederik Brooke Barnes}
\affiliation{Institute of Photonics and Quantum Sciences, Heriot-Watt University, Edinburgh EH14 4AS, United Kingdom}

\author{Moritz Cygorek}
\affiliation{Condensed Matter Theory, Technical University of Dortmund, 44227 Dortmund, Germany.}

\author{Antoine Borel}
\affiliation{Institute of Photonics and Quantum Sciences, Heriot-Watt University, Edinburgh EH14 4AS, United Kingdom}

\author{Natalia Herrera Valencia}
\affiliation{Institute of Photonics and Quantum Sciences, Heriot-Watt University, Edinburgh EH14 4AS, United Kingdom}



\author{Erik M. Gauger}
\affiliation{Institute of Photonics and Quantum Sciences, Heriot-Watt University, Edinburgh EH14 4AS, United Kingdom}

\author{Mehul Malik}
\affiliation{Institute of Photonics and Quantum Sciences, Heriot-Watt University, Edinburgh EH14 4AS, United Kingdom}

\author{Brian D. Gerardot}
\affiliation{Institute of Photonics and Quantum Sciences, Heriot-Watt University, Edinburgh EH14 4AS, United Kingdom}

\begin{abstract}
Quantum interference of indistinguishable photons is the foundation of photonic quantum technologies, yet scaling from a few to many identical quantum light sources remains a major challenge. In solid-state platforms, spatial and spectral inhomogeneity and resource-intensive architectures impede scaling. As a result, interference between remote, independent quantum emitters has been thus far limited to pairs. Here we introduce a wavefront-shaping approach that enables scalable interference from multiple indistinguishable quantum dots on the same chip. Using programmable spatial light modulators, we independently excite, collect, and route emission from spatially distinct, yet spectrally degenerate dots. Scaling from two to five indistinguishable emitters, we verify interference through cooperative-emission phenomena and Hong–Ou–Mandel two-photon interference, thereby establishing a route towards large-scale, programmable quantum photonic architectures.

\end{abstract}
 
\maketitle
 
\section{Introduction}

Solid-state quantum emitters can deliver near-unity single-photon purity, indistinguishability, and high brightness \cite{tomm2021microcavity, ding_brightqd_2025}, enabling major milestones in photonic quantum technologies. This includes high-rate quantum key distribution \cite{basset2021entangled_qkd, morrison_qkd_2023, yang_qkd_2025} and spin–photon interfaces that provide access to quantum memories for quantum networking and entangled cluster state generation \cite{Appel2025ManyBodyRegister,  cogan2023_clusterstate, Knaut2024TelecomNetworkEntanglement}. In addition, solid-state emitters can be embedded in monolithic or hybrid nanophotonic structures, forming deterministic light–matter interfaces that channel light directly into photonic circuitry \cite{Knaut2024TelecomNetworkEntanglement, Uppu2020ScalableIntegratedSPS, Larocque2024TunableFoundrySiliconPhotonics, Huang2025OnChipMoleculeInterference}. These attributes position solid-state quantum sources as building blocks for  quantum processors in which single photons generated from arrays of indistinguishable emitters are routed, interfered, and processed in a scalable fashion.

Despite this promise, scaling quantum interference between multiple independent and indistinguishable quantum light sources remains a central challenge. While high-visibility Hong-Ou-Mandel (HOM) interference has been achieved between photons from two emitters at either remote nodes \cite{zhai_quantum_2022} or on a single chip \cite{Huang2025OnChipMoleculeInterference}, extending quantum interference to larger numbers of independent solid-state emitters has remained out of reach. As a result, state-of-the-art demonstrations of entanglement generation have relied on active demultiplexing of photons from a single emitter~\cite{cao2024entangled,pont2024ghz,Maring2024VersatileSinglePhotonQC}, an approach that incurs substantial losses and resource overheads~\cite{Lenzini2017ActiveDemultiplexing}. These constraints fundamentally limit scalability and hinder the use of quantum emitters as practical engines for  quantum photonic architectures. 


Rather than fixed interferometric circuits, an alternative route to scalability is a programmable optical interface that selects, mode-matches, and coherently combines quantum light from multiple sources. This can be realized via wavefront shaping with programmable spatial light modulators (SLMs), which provide dynamic control over the transverse spatial structure of light. Wavefront-shaping techniques have been used for the control of quantum matter \cite{ebadi2021quantum-1ce,semeghini2021probing-cf9} and for programmable quantum photonic circuits \cite{goel_2025_qip,goel2024inverse-111}, but their integration with deterministic quantum light sources remains unexplored.

Here we introduce a wavefront-shaping approach that enables scalable quantum interference between multiple indistinguishable quantum dots (QDs) on a single chip. Using SLM-based wavefront shaping, we selectively excite and route photons in a reconfigurable manner from spectrally degenerate QDs at spatially distinct positions on the chip. This allows us to observe interference between up to five quantum emitters, confirming photon indistinguishability via cooperative-emission signatures and HOM measurements. Beyond this demonstration, our platform provides programmable photonic control, directly interfacing quantum light sources with reconfigurable photonic circuits and advancing the vision of quantum information processing with light.

\begin{figure*}[!ht]
    \centering
    \includegraphics[width = 0.95\textwidth]{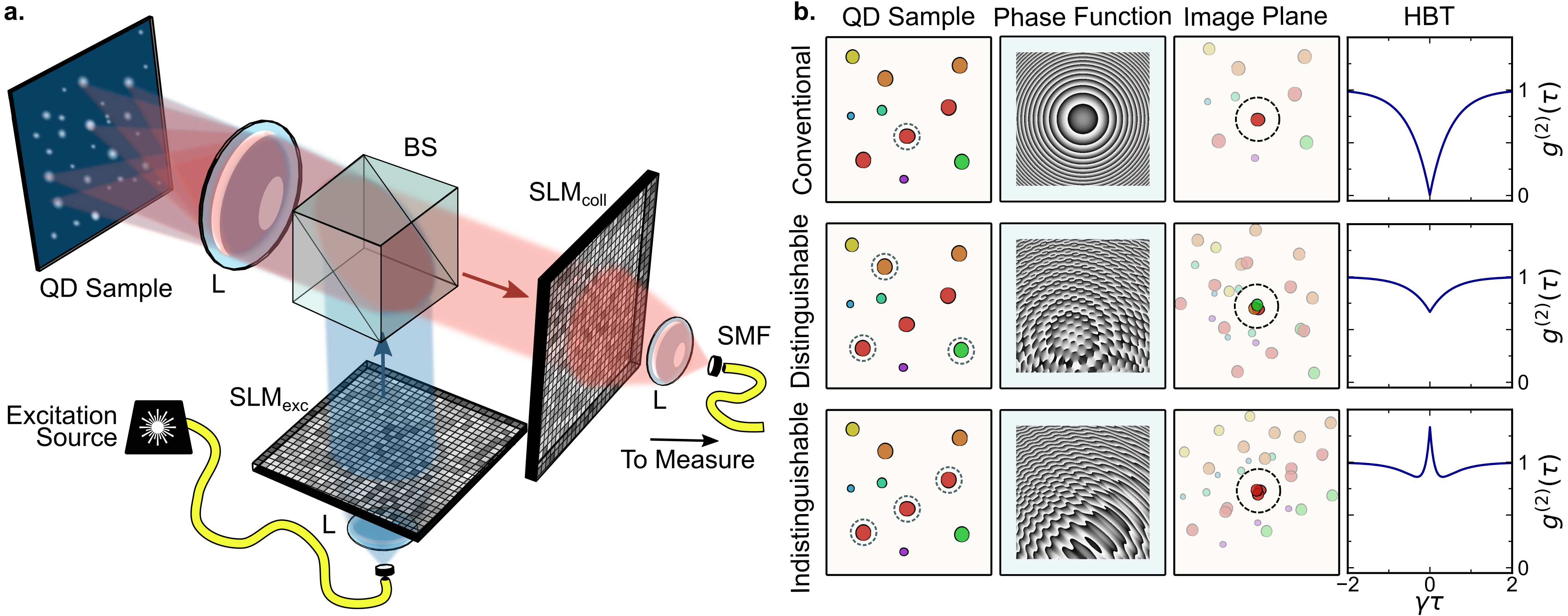}
    \caption{Programmable wavefront shaping and measurement-induced cooperative emission. \textbf{(a)} Experimental schematic. An above-band CW laser is wavefront-shaped by a digital hologram implemented on an excitation spatial light modulator ($\rm{SLM_{exc}}$) to focus on selected positions on the quantum dot (QD) sample. The emission is routed by the collection SLM ($\rm{SLM_{coll}}$) and a lens (L) into a common mode using a single-mode fiber (SMF) located at the image plane of the sample and sent to a spectrometer or an HBT interferometer for measurement. Abbreviations: BS, beamsplitter. \textbf{(b)} \textbf{Top row:} Schematic sketch of emission from a single QD mapped to the SMF collection mode by a phase function implemented by a combination of \(\mathrm{SLM}_{\mathrm{coll}}\) and a lens, yielding an HBT dip showing single-photon antibunching \(g^{(2)}(0)=0\). \textbf{Middle row:} Emission from three spatially separated, spectrally distinct QDs is superposed at a single collection mode using a multiplexed hologram on \(\mathrm{SLM}_{\mathrm{coll}}\). For \(N=3\) mutually distinguishable, equal-brightness sources, \(g^{(2)}(0)= 2/3\). \textbf{Bottom row:} Routing emission from three indistinguishable QDs to a common mode  erases which-path information and, upon detection, produces a zero-delay \emph{bunching} peak: \(g^{(2)}(0)> 1\).}
    
    \label{fig:fig1}
\end{figure*}

\begin{figure}[!htb]
    \centering
    \includegraphics[width = \linewidth]{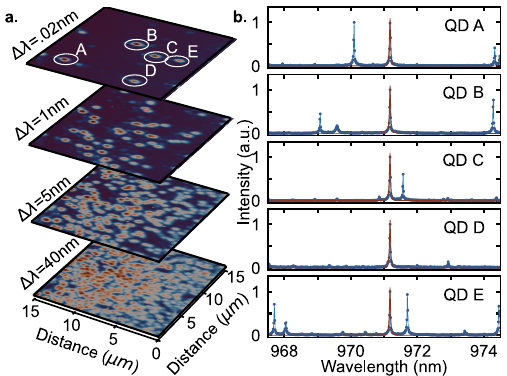}
    \caption{Spatial mapping to identify spectrally degenerate QDs.\textbf{(a)} Spatially resolved PL scans are performed over a $\rm{15 \times 15 \, \mu m}$ area on the sample by rastering the excitation spot with \(\mathrm{SLM}_{\mathrm{exc}}\) and routing emission with \(\mathrm{SLM}_{\mathrm{coll}}\) into a SMF and spectrometer. The spatial map is analyzed by post-selecting slices of narrow wavelength ranges to identify degenerate QDs. The uppermost slice reveals five spectrally degenerate QDs, labeled A, B, C, D and E, within a $\rm{0.02 \ nm}$ wavelength window centered at $\rm{971.17 \ nm}$.  \textbf{(b)} PL spectra of the $\rm{X^{1-}}$ transitions for QDs A--E. The blue curves are experimental data, the red curves Lorentzian fits for the selected QDs.}     
    
    \label{fig:fig2}
\end{figure}

\section{Measurement-induced cooperative emission via programmable wavefront shaping }

At the core of our approach are two SLMs that act as programmable optical interfaces: \(\mathrm{SLM_{exc}}\) sculpts the excitation wavefront to independently address chosen emitters and \(\mathrm{SLM_{coll}}\) sculpts the collection wavefront to route their photons into a single-mode fiber (SMF), as depicted in Fig.~\ref{fig:fig1}a. In the simplest configuration (Fig.~\ref{fig:fig1}b, top row), holograms displaying blazed gratings on both SLMs produce a diffraction-limited focus at the sample and at its image plane, as in a confocal microscope. Collected emission is directed to a spectrometer or a Hanbury Brown and Twiss (HBT) interferometer. Under continuous-wave excitation of a single QD, the second-order intensity correlation shows antibunching with \(g^{(2)}(0)\!\rightarrow\!0\) (Fig.~\ref{fig:fig1}b, top row).

To scale from one to many emitters, multiplexed holograms (a linear combination of phase patterns with variable amplitude weights; see Methods for details) on \(\mathrm{SLM_{exc}}\) generate multiple diffraction-limited foci at the sample, while a hologram on \(\mathrm{SLM_{coll}}\) imposes emitter-dependent phase offsets that bring the fields together at the image plane in a desired manner. This ensures that all selected emitters are routed into a common spatial mode at the fiber input and coupled to the SMF fundamental mode. 
For $N$ single-photon sources of equal brightness, the common-mode two-photon coincidences can be written as (see SI)
\begin{align}
    g^{(2)}(\tau) = 1-\frac{1}{N^2}\left(\mathcal{G}_\textrm{inc}(\tau) - \mathcal{G}_{\textrm{coh}}(\tau)\right), 
\label{eqn:eqn1}
\end{align}
where 
\begin{align}
    \mathcal{G}_\textrm{inc}(\tau) = \sum_{k=1}^{N}e^{-\tau(\gamma_{k}+\gamma_{k}^{p})}
\label{eqn:eqn2}
\end{align}
captures the individual contribution of each emitter and 
\begin{align}
    \mathcal{G}_{\textrm{coh}}(\tau) = \sum_{k,j=1}^{N} e^{-(i\Delta_{kj}+\Gamma_{kj})\tau}\,(1-\delta_{jk}),
\label{eqn:eqn3}
\end{align}
captures the impact of pairwise inter-emitter coherence. The independent-emitter contribution is only impacted by the incoherent decay and pumping rates \(\gamma_k\) and \(\gamma_k^{p}\) of the \(k^{\mathrm{th}}\) QD, respectively, while the inter-emitter coherences are determined by the energetic detuning between QDs \(k\) and \(j\), \(\hbar\Delta_{kj}\), and \(\Gamma_{kj}\), which sums all their decoherence contributions from incoherent driving, radiative decay, and pure dephasing.
If the emitted photons are clearly distinguishable, then the coherent contribution $\mathcal{G}_{\textrm{coh}}(\tau)$ averages to zero, leading to a zero delay-time value of $g^{(2)}(0) = 1 - \frac{1}{N}$. As an example, Fig.~\ref{fig:fig1}b (middle row) shows \(N=3\) spatially separated QDs emitting at different wavelengths mapped into a single superposed spatial mode (with all other modes rejected by the SMF), yielding \(g^{(2)}(0)=2/3\). 
On the contrary, if the photons emitted from the selected emitters are almost indistinguishable (e.g., matched in frequency, polarization, and temporal profile), then the two-photon coincidences retain the coherence contribution. This additional contribution results from the routing of the photons into a common mode, which erases which-path information and, upon detection, creates transient inter-emitter coherence. This effect has been termed `measurement-induced cooperative emission'~\cite{koong2022coopemission,cygorek_signatures_2023} and clearly demarcates itself from the emission from distinguishable emitters by the presence of a clear zero-delay bunching peak, with \(g^{(2)}(0)\) exceeding the distinguishable-source baseline \(1 - 1/N\) and further increasing with \(N\). This is sketched in Fig.~\ref{fig:fig1}b (bottom row) for $N=3$ QDs.

\section{Mapping, selecting, and Stark tuning of degenerate quantum dots}\label{sec:detuning_measurements}

We first map a relatively high- QD density sample ($\approx 6\ \mathrm{QDs}\,\mu\mathrm{m}^{-2}$), chosen to increase the statistical likelihood of finding spectrally degenerate emitters. The device is charge-tunable (see Methods for details), and we operate on the negatively charged exciton (\(X^{-}\)) plateau to ensure a stable QD charge state and linewidth. Using \(\mathrm{SLM}_{\mathrm{exc}}\) to raster-scan (0.15 $\mu m$ step size) the above-band excitation laser and \(\mathrm{SLM}_{\mathrm{coll}}\) to route emission into a single-mode fiber (SMF), we record a spectrum at every pixel and post-select narrow wavelength slices to reveal degenerate emitters (Fig.~\ref{fig:fig2}a). A 0.02\,nm window centered at 971.17\,nm identifies five spatially separated QDs (A--E) with a common \(X^{-}\) transition energy (within the spectrometer resolution, \(\approx 30\,\mu\mathrm{eV}\); Fig.~\ref{fig:fig2}b).

\begin{figure}[!htb]
    \centering
    \includegraphics[width = \linewidth]{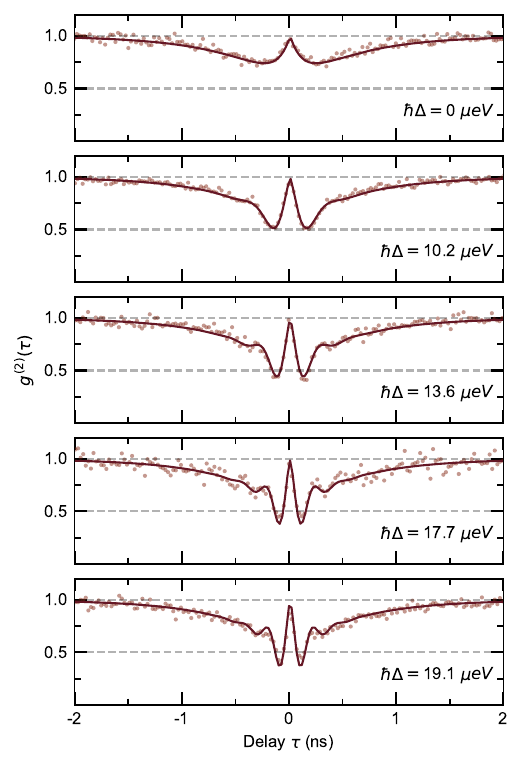}
    \caption{Second-order correlation measurements of cooperative emission for QDs A and B at selected detuning ($\Delta$) values. At resonance (\(\Delta = 0\)) a zero-delay bunching peak is observed. As $\Delta$ increases, the central feature narrows and beat notes emerge at frequency $\Delta$, consistent with two-emitter quantum interference. Data points represent raw data and the  curves are IRF-convolved fits to a two-emitter model.}
    
    \label{fig:fig3}
\end{figure}

From this set of QDs we select dots A and B, which exhibit different permanent dipoles that enable relative DC Stark tuning of their optical transitions~\cite{koong2022coopemission}. With $\mathrm{SLM}_{\mathrm{exc}}$ programmed to excite both dots and $\mathrm{SLM}_{\mathrm{coll}}$ to coherently combine their emission with equal amplitude into a common SMF mode, we sweep the gate voltage to control the inter-dot detuning \(\Delta\) and drive the pair into and out of resonance. The corresponding $\rm{g^{(2)}(\tau)}$ traces (Fig.~\ref{fig:fig3}) exhibit zero-delay bunching at resonance and a narrowing central peak with emerging beat notes as \(\Delta\) increases, consistent with quantum-interference effects. We fit the experimental data to the analytical equation (Eqn.~\ref{eqn:eqn1}) convolved with a Gaussian instrument response function with a full width at half maximum (FWHM) of $\rm{35 \, ps}$. Considering $\gamma \, + \gamma_{p}$ as a fixed fitting parameter obtained from the single dot $\rm{g^{(2)}(\tau)}$ (see Methods), we extract a single, constant dephasing rate ($\gamma_d$) $\rm{\approx \, 3.0 \, ns^{-1}}$ for all $\Delta$. We calculate $\Delta$ from the beating frequency of the fit and obtain $\hbar\Delta$ values ranging from $\rm{0 \, \mu eV}$ to $\rm{19.1 \, \mu eV}$, as labelled in Fig.~\ref{fig:fig3}.

 

\begin{figure*}[t]
    \centering
    \includegraphics[width = \textwidth]{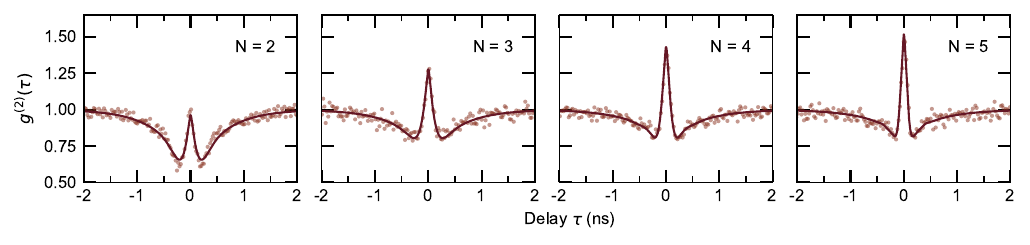}
    \caption{Second-order intensity correlations \(g^{(2)}(\tau)\) for \(N=2,3,4,5\) indistinguishable emitters routed into the SMF fundamental mode reveals zero-delay bunching peaks increasing with \(N\): \(g^{(2)}(0)=0.96\pm 0.027,\,1.28 \pm 0.028,\,1.43 \pm 0.025,\,1.52 \pm 0.029 \) for \(N=2,3,4,5\), respectively. Points: data; red curves: analytical fits.}
    \label{fig:fig4}
\end{figure*}

\section{Programmable multi-emitter interference}

We next scale quantum interference by routing indistinguishable photons from subsets of spectrally degenerate emitters (A--E in Fig.~\ref{fig:fig2}) with equal brightness into the SMF fundamental mode, increasing \(N\) sequentially from 2 to 5. The resulting \(g^{(2)}(\tau)\) traces (Fig.~\ref{fig:fig4}) show a pronounced zero-delay bunching peak that grows with \(N\): \(g^{(2)}(0)=0.96 \pm 0.027,\,1.28 \pm 0.028,\,1.43 \pm 0.025,\) and \(1.52 \pm 0.029\) for \(N=2,3,4,\) and 5, respectively. Each dataset is well described by analytic fits (red curves) to Eq.~(\ref{eqn:eqn1}), convolved with a Gaussian instrument response function (FWHM \(\sim\) 35\,ps). As in the two-QD detuning analysis, we fix \(\gamma+\gamma_{p}\) from single-emitter \(g^{(2)}(\tau)\) and find that a single dephasing rate \(\gamma_{d}\approx 3.0\,\mathrm{ns}^{-1}\) describes all datasets, consistent with the values in Sec.~\ref{sec:detuning_measurements}. All measured \(g^{(2)}(0)\) values exceed the distinguishable-source baseline \(1-1/N\) (\(0.50,\,0.67,\,0.75,\,0.80\) for \(N=2,3,4,\) and 5), demonstrating measurement-induced cooperativity arising from indistinguishable photons routed into a common spatial mode. Small deviations from the ideal trend arise from finite dephasing, imperfect brightness balancing, and residual pairwise detunings \(\Delta_{kj}\) (see Supplementary Information).

\begin{figure}
    \centering
   
        \includegraphics[width = \linewidth]{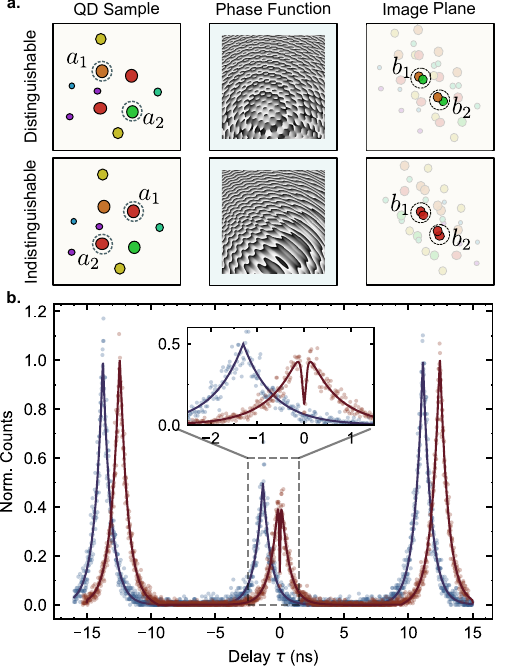}
    \caption{Pulsed HOM measurement on two QDs. (a) Conceptual sketch of the experiment. Two QDs are selected and the hologram on the SLM maps the distinct spatial modes with annihilation operators $a_1$ and $a_2$ onto superposition modes $b_1$ and $b_2$, such that $b_1= a_1+a_2$ and $b_2 = a_1-a_2$, which are collected by two separate cores of a multi-core fiber and directed to independent single photon detectors for corelation measurements. (b) HOM results for the two QDs in resonance (red) and out of resonance (blue). Theory fits are shown by solid lines. The results out of resonance are shifted by $1.5$~ns for visual clarity.}
    \label{fig:fig5}
\end{figure}

Beyond many-emitter cooperative emission, HOM interference---manifested as a suppression of coincidence  events when two indistinguishable photons interfere at a balanced beamsplitter---provides a stringent benchmark of pairwise photon indistinguishability. We therefore demonstrate that the same wavefront-shaping platform can be reconfigured to implement an effective beamsplitter and enable HOM interference~\cite{leedumrongwatthanakun2020programmable}. Unlike the preceding HBT measurements, HOM interference requires two distinct spatial output modes corresponding to the two ports of a balanced beamsplitter.
We implement this beamsplitter transformation on \(\mathrm{SLM_{coll}}\) via an inverse-design algorithm known as wavefront matching ~\cite{sakamaki_new_2007,hashimoto_optical_2005}. This algorithm uses the overlap between the spatial modes corresponding to the emission from the two selected QDs, and the output modes corresponding to two cores of a multi-core fiber (MCF), to calculate a phase function (see Fig.~\ref{fig:fig5}\textbf{(a)}). This phase function is implemented on $\mathrm{SLM_{coll}}$ and maps the emission from two selected quantum dots onto two superposition modes, effectively functioning as a 50{:}50 beamsplitter.
The two output modes are collected into independent cores of an MCF and directed to the detectors, enabling coincidence measurements between the beamsplitter outputs.

As a control, we first select a pair of distinguishable QDs (\(\hbar\Delta = 32\,\mu\mathrm{eV}\)) separated by \(\approx 7\,\mu\mathrm{m}\). The emitters are simultaneously excited using pulsed phonon-assisted excitation~\cite{Reindl2019HighlyIndistinguishablePRB}. The resulting second-order correlation function yields \(g^{(2)}(0) = 0.50 \pm 0.04\) (blue points in Fig.~\ref{fig:fig5}\textbf{(b)}), as expected for distinguishable photons at the output of a balanced beamsplitter, thereby validating the beamsplitter operation. We next select two spectrally degenerate QDs (separation \(\approx 11\,\mu\mathrm{m}\)). Under pulsed phonon-assisted excitation, we observe a pronounced suppression of coincidence events at \(\tau=0\) (red points in Fig.~\ref{fig:fig5}\textbf{(b)}). Fitting a theoretical model to the data, we extract \(g^{(2)}(0) = 0.13 \pm 0.04\), evidencing Hong–Ou–Mandel interference between the two emitters.

To quantify two-photon interference, we define the HOM visibility from the zero-delay coincidence as 
\(\mathcal{V}_0 = 1 - g^{(2)}_{\mathrm{ind}}(0)/g^{(2)}_{\mathrm{dis}}(0)\), 
where \(g^{(2)}_{\mathrm{ind}}(0)\) and \(g^{(2)}_{\mathrm{dis}}(0)\) correspond to indistinguishable and distinguishable photons, respectively. Using this definition, we obtain \(\mathcal{V}_0 \approx 73.7 \pm 8.4\%\). The measured visibility is primarily limited by the finite coherence time of the QDs, which lack Purcell enhancement, as well as the phonon-assisted preparation scheme, rather than by imperfections in the SLM-based interferometer.

\section{Outlook}

We have introduced a wavefront-shaping approach that enables scalable quantum interference from multiple indistinguishable quantum emitters on a single chip. Using programmable SLMs, we independently excite, collect, and interfere single-photon emission from up to five spatially distinct emitters. Within the same platform, we reconfigure the optical mapping to realize either single-mode output, yielding cooperative-emission signatures, or a two-output beam splitter for Hong–Ou–Mandel interference. These results establish wavefront shaping as a flexible and scalable interface between solid-state quantum emitters and programmable photonic architectures, overcoming key constraints imposed by spatial inhomogeneity and fixed interferometric circuitry.


Looking ahead, the full potential of this approach lies in extending wavefront shaping to reconfigurable optical circuits for spatially structured light \cite{goel_2025_qip}. The use of transverse-spatial modes allows one to break out of the limited planar architecture of photonic integrated circuits, significantly increasing scalability. Spatial-mode circuits have been demonstrated with many platforms including multi-mode fibres \cite{goel2024inverse-111, Valencia_2026}, multi-plane light converters \cite{Brandt_2020,Lib_2025}, and multi-core fibres \cite{Lima_2023}. However, implementation challenges remain in terms of circuit scalability and loss. Nevertheless, combining deterministic indistinguishable solid-state emitters and efficient nanophotonic interfaces with wavefront shaping unites high-quality sources and arbitrary photonic transformations within a single platform, offering a scalable path toward universal linear-optical networks and programmable quantum photonic processors.

\begin{acknowledgements}
 This work was made possible by financial support from the 
Leverhulme Trust through grant number RPG-2022-335, EIC Pathfinder Challenge 101161312 through Innovate UK Grant No.~10120741, the 
Engineering and Physical Sciences Research Council (EPSRC) (EP/Z533166/1 and EP/Z533208/1), European Research Council (ERC) Starting Grant PIQUaNT (950402), and the Royal Academy of Engineering Chair in Emerging Technologies program.
\end{acknowledgements}



\bibliography{bib}

\section*{Methods}
\subsection{Theoretical model} 
For our theoretical model, we approximate each QD as a two-level system (TLS) with ground state $|g_k\rangle$ and excited state $\ket{e_k}$, $k = 1, \dots 5$. Each QD is described by its transition energy, $\hbar\omega_k$, its decay rate $\gamma_k$ and its incoherent pumping and pure dephasing rates $\gamma_k^p$ and $\gamma_k^d$, respectively. Before the first photon detection event produces inter-emitter coherence the QDs, evolve independently and the state $\rho(t)$ of the system therefore is a product state
\begin{equation}
    \rho(t) = \prod_{k=1}^N \rho_k(t).
\end{equation}
The dynamics of each QD can be obtained by solving the Lindblad master equation for each QD independently:
\begin{align}
    \frac{d}{dt}\rho_k = &-i\omega_k[\sigma_{k}^+\sigma_k^-, \rho_k] + \gamma_k\mathcal{L}_{\sigma_k^-}[\rho_k] \\
    &+ \gamma_k^p\mathcal{L}_{\sigma_k^+}[\rho_k] + \gamma_k^d\mathcal{L}_{\sigma_k^+\sigma_k^-}[\rho_k],
\end{align}
where $\sigma_k^+ = \ket{e_k}\bra{g_k}$ and $\sigma_k^- = \ket{g_k}\bra{e_k}$, and
\begin{align}
    \mathcal{L}_{O}[\rho] = O\rho O^\dagger - \frac{1}{2}[O^\dagger O \rho + \rho O^\dagger O].
\end{align}
are Lindblad operators \cite{breuer_theory_2009}. 
We then obtain the system steady state as
\begin{align}
    {\rho}_{SS} = \prod_{k=1}^{N}{\rho}_{SS}^{k} = \prod_{k=1}^{N} \left\{\ket{e_k}\bra{e_k}n_k^e + \ket{g_k}\bra{g_k}n_k^g\right\}.
\end{align}
Here, we have introduced the excited (ground) state occupation $n_k^e$ ($n_k^g$) of the individual QD $k$. 

We describe the action of the SLM in the experiment assuming that it maps sets of local electromagnetic modes around each QD, described by the creation operator $a^\dagger_{k, \alpha}$, where $k$ indicates the QD and $\alpha$ is the local mode index to a collective mode. We then write the creation operator of the collective mode after the SLM as
\begin{align}
    b^\dagger = \sum_{k = 1}^{N}\sum_{\alpha}\beta_{k,\alpha}a^\dagger_{k, \alpha}
\end{align}
with some complex coefficients $\beta_{k, \alpha}$, which are determined by the programming of the SLM. Using the Heisenberg equations of motion, the intensity recorded at the detector after the SLM can be written as
\begin{align}
    I(t) = \langle{b^\dagger}b\rangle(t) = \langle \sigma_C^+\sigma_C^-\rangle(t), 
\end{align}
with the collective transition operator $\sigma_C^+ = \sum_{k=1}^N\sum_{\alpha}\beta_{k, \alpha}\sigma_{k}^{+}$. Now, we assume that the SLM is adjusted such that each QD, irrespective of its brightness, contributes equally,  that is $N|\sum_\alpha\beta_{k, \alpha}|^2n_{k}^e = I_0$. Then, the HBT signal can be calculated to be (see also SM \cite{SM})
\begin{align}
    g^{(2)}&(\tau) = \lim_{t\to\infty} \frac{\langle\sigma_C^+(t)\sigma_C^+(t+\tau)\sigma_C^-(t+\tau)\sigma_C^{-}(t)\rangle}{I_0^2}\nonumber\\
    &=1 - \frac{1}{N^2}\sum_{k= 1}^N\left(e^{-\tau(\gamma_k+\gamma_k^p)})- \sum_{\substack{j=1\\k\neq j}}^Ne^{-(i\Delta_{kj} + \Gamma_{kj})\tau}\right)\label{eq:met_g2}, 
\end{align}
where $\Gamma_{kj} = (\gamma_k + \gamma_j + \gamma^p_k + \gamma^p_j + \gamma^d_k + \gamma^d_j )/2$ is the sum of the decoherence contributions from two QDs $k$ and $j$, and $\Delta_{kj} = \omega_k-\omega_j$ is their respective detuning. From this equation the bounds for the maximum value of $g^{(2)}(0)$ can be determined. As outlined in the supplementary material, the impact of phonon effects on the anti-dip height, which result from emission into the phonon sidebands can be estimated via polaron theory \cite{wiercinski_phonon_2023,iles-smith_phonon_2017}

We fit the experimental data using a reduced $\chi^2$ method that allows fitting a set of parameters consistently across multiple experiments \cite{quach_superabsorption_2022, hallett_controlling_2026}. This allows predicting a common dephasing rate describing all data presented in Fig.~\ref{fig:fig3}, while estimating the detuning dependence on the bias voltage. Similarly, it enables us to find a set of consistent dephasing rates and detunings between multiple quantum dots that describe the data presented in Fig.~\ref{fig:fig4}. Additional information on the fitting procedure, as well as the resulting parameters and their error estimates, is given in the Supplementary Material \cite{SM}.

\subsection{Sample and experimental setup details}
The sample consists of self-assembled InGaAs QDs in a GaAs membrane with a bottom gold mirror and top solid-immersion lens  to enhance photon-collection efficiency \cite{koong2022coopemission}. A Schottky diode is used to control the  QD charge state via Coulomb blockade and fine-tune the QD energy via DC Stark tuning. We use $\lambda$ = 830 nm continuous-wave (CW) laser excitation for QD mapping and cooperative emission measurements (Figs 2 - 4) and a tunable Ti:Saphire laser with $\approx$ 20 ps pulse duration for for HOM interference measurements (Fig. 5). Emission from the QDs is analysed either with a spectrometer ($\approx$ 30 $\mu$eV resolution) or via photon-correlation measurements in an HBT/HOM configuration using superconducting nanowire single-photon detectors ($\approx$ 35 ps time jitter). To suppress residual laser leakage, PL from other QDs, and incoherent phonon sideband contributions to the PL, we use a long-pass  ($\lambda$ = 950 nm) filter and a grating-based spectral filter ($\approx$ 100 $\mu$eV resolution).

\subsection{Generating holograms to map and measure multiple emitters}
In order to excite (or redirect) light from (or to) various different accessible locations on the sample, we display computer-generated holograms on $\rm{SLM_{exc}}$ ($\rm{SLM_{coll}}$).
Since the SLMs lie at the back-focal plane of the sample, the spatial field corresponding to each location on the sample $r_k$ is approximated as a  plane wave $\ket{f_k}$ calculated by assuming a Fourier transfer function due to a lens with an effective focal length of the three-lens system between each SLM and the sample~\cite{goodman2005introduction}.
Therefore, to excite a QD located at position $r_k$, the $\rm{SLM_{exc}}$ displays a hologram that redirects the incident laser light into the spatial mode $f_k^{\text{exc}}$. 
Likewise, the emission from a location $r_k$ on the sample is collected into an SMF by displaying a hologram corresponding to the spatial mode $f_k^{\text{coll}}$ on $\rm{SLM}_{coll}$.
The holograms displayed on $\rm{SLM_{exc}}$ and $\rm{SLM_{coll}}$ are calculated by taking the phase component of the corresponding fields, i.e.  $\phase{f_k^{\text{exc}}}$ and $\phase{f_k^{\text{coll}}}$ respectively, and adding them to pre-defined carrier gratings in order to disregard the imperfections of the SLMs~\cite{arrizon2007pixelated}. 

To address multiple QDs located at positions $\{ r_k\}_{k=1}^N$ at the same time, a corresponding optical field on $\rm{SLM_{exc}}$ that excites them together is calculated as $\tilde{f}^{\text{exc}} = \sum_{k = i}^N \alpha_k\ket{f^{\text{exc}}_k}$
where $\alpha_k$ are the amplitude weights with which each QD is excited.
Similarly, to collect emission from multiple locations, optical field on $\rm{SLM_{coll}}$ is calculated as $\tilde{f}^{\text{coll}} = \sum_{k = i}^N \beta_{k}\ket{f^{\text{coll}}_k}$
where $\beta_k$ are the amplitude weights with which emission from each QD is collected. 
Once again, the holograms displayed on each SLM are generated by taking the phase component of the respective fields , i.e. $\phase{\tilde{f}^{\text{exc}}}$ and $\phase{\tilde{f}^{\text{coll}}}$, and adding respective pre-defined carrier gratings. 

Finally, to perform HOM measurements, emission is collected from two QDs located at $r_1$ and $r_2$ on the sample, and coupled into two single-mode cores located at $r^{\text{SMF}}_1$ and $r^{\text{SMF}}_2$ of a multi-core fiber, after performing a beam splitter operation. 
To achieve this, $\rm{SLM}_{coll}$ needs to map transformed spatial modes corresponding to emitted light, i.e. $\ket{f^{\text{coll}}_1} + \ket{f^{\text{coll}}_2}$ and $\ket{f^{\text{coll}}_1} - \ket{f^{\text{coll}}_2}$, to that of the spatial modes corresponding to each SMF, which we denote by $\ket{f^{\text{SMF}}_1}$ and $\ket{f^{\text{SMF}}_2}$.
The corresponding collection phase-pattern is generated using a technique similar to the wavefront matching algorithm~\cite{hashimoto_optical_2005, sakamaki_new_2007}, where only a single phase-layer is used between the optical modes.
The final collection hologram is implemented with the addition of the bespoke collection carrier grating.

\clearpage
\onecolumngrid
\appendix

\end{document}